\documentclass[prl,twocolumn,showpacs]{revtex4}

\usepackage{amsmath}
\usepackage{graphicx}

\begin{document}

\title{Turing patterns on networks}

\author{Hiroya Nakao$^{1,2}$ and Alexander S. Mikhailov$^{2}$}

\affiliation{$^{1}$Department~of~Physics,~Kyoto~University,~Kyoto~606-8502,~Japan}

\affiliation{$^{2}$Abteilung~Physikalische~Chemie,
  Fritz-Haber-Institut~der~Max-Planck-Gesellschaft,
  Faradayweg~4-6,~14195~Berlin,~Germany}

\begin{abstract}
  Turing patterns formed by activator-inhibitor systems on networks
  are considered.  The linear stability analysis shows that the Turing
  instability generally occurs when the inhibitor diffuses 
  sufficiently faster than the activator.
  Numerical simulations, using a prey-predator model on a scale-free
  random network, demonstrate that the final, asymptotically reached
  Turing patterns can be largely different from the critical
  modes at the onset of instability, and multistability and
  hysteresis are typically observed.
  An approximate mean-field theory of nonlinear Turing patterns on the
  networks is constructed.
\end{abstract}
\date{\today}

\maketitle

Turing instability in activator-inhibitor systems is one of the most
important classical concepts of nonequilibrium pattern
formation~\cite{Turing}, with diffusion destabilizing a uniform
stationary state of the system and leading to formation of stationary
spatial patterns. Turing patterns in ordinary continuous media have been
extensively investigated in theoretical~\cite{Walgraef} and
experimental~\cite{DeKepper,Ouyang} studies. Their
properties have also been studied for spatially heterogeneous
systems~\cite{Page} and in the presence of fractional
diffusion~\cite{Henry}.

In this Letter, we analyze Turing patterns exhibited by
activator-inhibitor systems on networks. In such systems, the nodes
are populated by activator and inhibitor species which are diffusively
transported over the edges that form a network. As examples,
multicellular systems~\cite{Othmer}, networks of coupled chemical
reactors~\cite{Horsthemke}, and transportation networks that
facilitate spreading of animals or insects~\cite{Urban, Fortuna} can
be mentioned. If several species are present, their mobilities on a
network may significantly differ, and a situation is possible where
the inhibitor species is much more easily transported over a network
than the activator. Previously, it has been shown by linear stability
analysis that the uniform stationary state in such networks can be
unstable~\cite{Othmer, Horsthemke}.  However, detailed numerical and
analytical investigations have so far been performed only for
lattices~\cite{Othmer} and for small network systems~\cite{Horsthemke}.

Here, we provide a complete linear stability analysis valid for any
activator-inhibitor system and perform a detailed numerical study of
Turing patterns in the Mimura-Murray model of interacting
prey-predator populations~\cite{Murray}.  Our numerical results reveal
that the actual nonlinear behavior of network-based
activator-inhibitor systems may show significant differences from the
predictions of the linear stability analysis, with coexistence of many
different structures and hysteresis being typically observed. As we
further show, the properties of network Turing patterns can often be
well understood in the framework of a mean-field approximation.

Activator-inhibitor systems on networks are described by equations
\begin{align}
  \dot{U_{i}}(t) & =f(U_{i},V_{i})+\epsilon(\nabla^{2} U)_{i},\cr
  \dot{V_{i} }(t) & =g(U_{i},V_{i})+\sigma\epsilon(\nabla^{2}
  V)_{i},\label{Eq:RD}
\end{align}
where $U_{i}$ and $V_{i}$ are concentrations of the activator and the
inhibitor species on node $i=1,\cdots, N$, and functions $f(U,V)$ and
$g(U,V)$ determine their local dynamics on a single node. In these
equations, $\nabla^{2}$ represents the diffusion (or Laplacian)
operator on the network defined as $(\nabla^{2} U)_{i}=\sum_{j=1}^{N}
A_{ij} (U_{j}-U_{i})$ and $(\nabla^{2} V)_{i}=\sum_{j=1}^{N} A_{ij}
(V_{j}-V_{i})$, where $(A_{ij})$ is the adjacency matrix of the
network.  All nodes and edges in the network have the same
properties. The parameter $\epsilon$ is the diffusion constant of the
activator, and $\sigma$ is the ratio of the inhibitor diffusion
constant to that of the activator. Note that the action of the network
diffusion operator can also be expressed as $(\nabla^{2}
U)_{i}=\sum_{j=1}^{N} L_{ij}U_{j}$ where $(L_{ij})$ is the Laplacian
matrix of the network, defined as $L_{ij} =A_{ij}-k_{i}\delta_{ij}$
with $k_{i}=\sum_{j=1}^{N}A_{ij}$ representing the degree of the node
$i$.

In absence of diffusion ($\epsilon=0$), each network element has a
linearly stable stationary state $(U^{(0)},V^{(0)})$, which satisfies
$f(U^{(0)}, V^{(0)})=0$ and $g(U^{(0)},V^{(0)})=0$. Since $U$ is the
activator and $V$ is the inhibitor, we assume $f_{u}=\left.  \partial
  f/\partial U\right\vert _{(U^{(0)},V^{(0)})}>0$,
$f_{v}=\left.  \partial f/\partial V\right\vert
_{(U^{(0)},V^{(0)})}<0$, $g_{u}=\left.  \partial g/\partial
  U\right\vert _{(U^{(0)},V^{(0)})}>0$, and $g_{v}=\left.  \partial
  g/\partial V\right\vert _{(U^{(0)},V^{(0)})}<0$. When diffusion is
turned on ($\epsilon>0$), the uniform solution
$(U_{i},V_{i})\equiv(U^{(0)},V^{(0)})$ still satisfies
Eqs.~(\ref{Eq:RD}). However, this uniform solution may now become
unstable due to the difference in diffusional mobilities of the
activator and inhibitor species.

\begin{figure}[ptbh]
\begin{center}
\includegraphics[width=0.7\hsize]{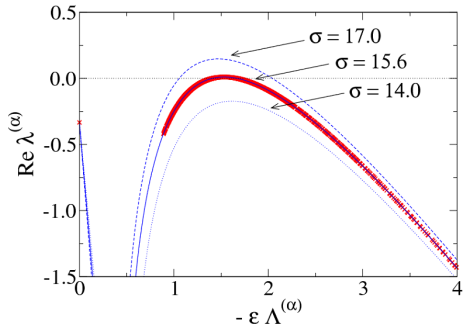}
\end{center}
\caption{(color online) Linear growth rates in the Mimura-Murray model
  on the scale-free network, with $\mbox{Re}\ \lambda^{(\alpha)}$
  plotted against $-\epsilon\Lambda^{(\alpha)}$ for $\epsilon=0.12$
  and three different values of the ratio $\sigma$ of the diffusion
  constants. Crosses show actual discrete values of the growth rates
  for $\sigma=15.6$. }
\label{Fig1}
\end{figure}

The linear stability analysis is naturally performed by using
eigenvalues $\Lambda^{(\alpha)}$ and eigenvectors
$\phi^{(\alpha)}=(\phi _{1}^{(\alpha)},\cdots,\phi_{N}^{(\alpha)})$ of
the Laplacian matrix, satisfying $(\nabla^{2}
\phi^{(\alpha)})_{i}=\sum_{j=1}^{N}L_{ij}\phi_{j}^{(\alpha
  )}=\Lambda^{(\alpha)}\phi_{i}^{(\alpha)}$ ($\alpha= 1, \cdots,
N$). Since $L_{ij}$ is a real symmetric matrix, all its eigenvalues
are real, and its eigenvectors are mutually orthogonal. We sort
indices $\{\alpha\}$ in the decreasing order of the eigenvalues, so
that $0=\Lambda ^{(1)}>\Lambda^{(2)}>\cdots>\Lambda^{(N)}$ holds.

Linearized equations describing evolution of small perturbations
$(u_{i},v_{i})$ to the uniform state $(U^{(0)},V^{(0)})$ are given by
\begin{align}
  \dot{u}_{i}(t) & = f_{u}u_{i}+f_{v}v_{i}+\epsilon(\nabla^{2} u)_{i},
  \cr \dot{v}_{i}(t) & =
  g_{u}u_{i}+g_{v}v_{i}+\sigma\epsilon(\nabla^{2}
  v)_{i}.\label{Eq:LRD}
\end{align}
By expanding $(u_{i}, v_{i})$ over the Laplacian eigenvectors,
$(u_{i}, v_{i}) = \sum_{\alpha=1}^{N} (u^{(\alpha)}, v^{(\alpha)})
\exp \left[ \lambda^{(\alpha)}t \right] \phi_{i}^{(\alpha)}$ where
$u^{(\alpha)}$ and $v^{(\alpha)}$ are the expansion coefficients,
Eqs.~(\ref{Eq:LRD}) are transformed into $N$ independent linear
equations for different modes $(u^{(\alpha)},v^{(\alpha)})$. The
linear growth rate $\lambda^{(\alpha)}$ of the $\alpha$-th mode is
determined from the characteristic equation
$\{\lambda^{(\alpha)}-f_{u}-\epsilon\Lambda^{(\alpha)}\}\{\lambda^{(\alpha
  )}-g_{v}-\sigma\epsilon\Lambda^{(\alpha)}\}-f_{v}g_{u}=0$, which
yields
\begin{align}
  \lambda^{(\alpha)} & = \left( f_{u}+g_{v}+(1+\sigma)\epsilon\Lambda
    ^{(\alpha)} \right.  \cr & \left.  \pm\sqrt{4f_{v}g_{u}+\left(
        f_{u}-g_{v} +(1-\sigma)\epsilon\Lambda^{(\alpha)}\right) ^{2}}
  \right) / \;2.\label{growthrate}
\end{align}
When $\mbox{Re}\ \lambda^{(\alpha)}$ is positive, the $\alpha$-th mode
is unstable. The instability takes place when one of the modes begins
to grow, namely, when $\mbox{Re }\lambda^{(\alpha)} = 0$ for some
$\alpha = \alpha_{0}$ and $\mbox{Re }\lambda^{(\alpha)} < 0$ for all
other modes.  For the Turing instability, the condition $\mbox{Im}\
\lambda^{(\alpha)} = 0$ should also hold.

Graphically, all growth rates $\lambda^{(\alpha)}$ lie on the curve
$\lambda = F(\Lambda)$, which is determined by Eq.~(\ref{growthrate})
(see Fig.~\ref{Fig1}). When the ratio $\sigma$ of the diffusion
constants is varied, this curve touches the horizontal axis at
$\sigma_{c}=\left( f_{u} g_{v}-2f_{v}g_{u} +2\sqrt{f_{v}g_{u}\left(
      f_{v}g_{u}-f_{u}g_{v}\right) }\right) \;/\;f_{u}^{2}$, and some
of the modes may therefore become unstable for $\sigma >
\sigma_{c}$. For this to occur, the eigenvalues of the discrete
Laplacian spectrum of the considered network must be actually present
near the maximum of the curve.  Therefore, the condition
$\sigma=\sigma_{c}$ provides only the lower bound for the Turing
instability boundary.

Thus, the condition of the Turing instability on networks is similar
to the respective condition for the continuous media.  The role of
plane waves is now played by Laplacian eigenvectors of the considered
network, and, instead of wavenumbers, the respective Laplacian
eigenvalues are important.  Note that the above linear stability
analysis is general; it holds for any activator-inhibitor model and
for any network architecture. The critical mode always corresponds to
a certain Laplacian eigenvector.

\begin{figure}[ptbh]
\begin{center}
\includegraphics[width=1.0\hsize]{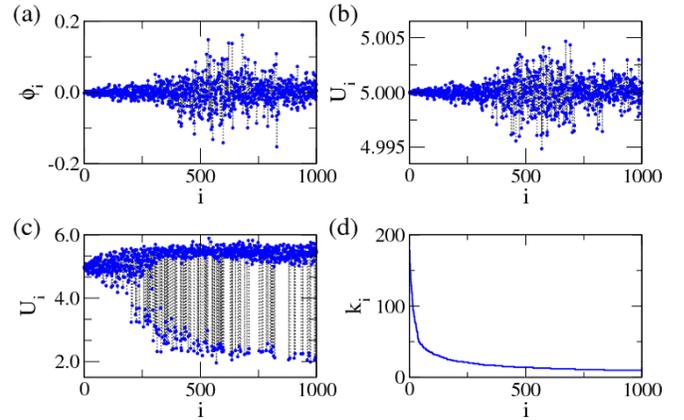}
\end{center}
\caption{(color online) (a) The critical mode (the Laplacian
  eigenvector with $\alpha_0=422$), (b) the activator pattern at the
  early evolution stage ($t=200$), and (c) the stationary activator
  pattern at the late stage ($t=1500$).  Nodes are ordered according
  to their degrees; with (d) showing the dependence of the degree on
  the node index. }
\label{Fig2}
\end{figure}

To perform numerical investigations of nonlinear dynamics, a specific
activator-inhibitor model is needed. For this purpose, we have chosen
the Mimura-Murray model~\cite{Murray}, which has been introduced to
explain spatial nonuniformity of prey-predator populations in
ecological systems.  The chosen model is given by the equations
$f(U,V)=\{(35+16U-U^{2})/9-V\}U$ and $g(U,V)=\{U-(1+2V/5)\}V$, whose
fixed point is $(U^{(0)},V^{(0)})=(5,10)$. As an example of random
networks, we take below the Barab\'{a}si-Albert scale-free
networks~\cite{Barabasi} of size $N=1000$. They are generated by the
preferential attachment algorithm, starting from $10$ fully connected
initial nodes and adding $m=10$ new connections at each iteration
step, so that their mean degree is $\langle k\rangle \simeq 2 m = 20$.

Figure~\ref{Fig1} displays the real part of the linear growth rate
$\mbox{Re}\ \lambda^{(\alpha)}$ for three values of $\sigma$ with
$\epsilon=0.12$.  The growth rate can become positive for $\sigma >
\sigma_{c}=15.5$. The largest growth rate is reached at
$\alpha_{0}=422$ for $\sigma=15.6$.  Near $\alpha = \alpha_0$,
$\mbox{Im}\ \lambda^{(\alpha)} = 0$ holds.
In Fig.~\ref{Fig2}, we compare the critical mode ($\alpha_{0}=422$)
with two snapshots of the actual activator patterns at the early and
late stages of the evolution, as yielded by numerical integration of
Eqs.~(\ref{Eq:RD}) with $\epsilon=0.12$ and $\sigma=15.6$.
Here and below, node indices $\{i\}$ are sorted in the decreasing
order of their node degrees $\{k_{i}\}$ so that $k_{1}\geq
k_{2}\geq\cdots\geq k_{N}$ holds, which is useful in visualizing the
Turing patterns exhibited by our system.

Starting from almost uniform initial conditions with tiny
perturbations, exponential growth is observed at the early stage. The
activator pattern at this stage, Fig.~\ref{Fig2}(b), is similar to the
critical mode, Fig.~\ref{Fig2}(a), with the deviations due to
contributions from a few neighboring modes that are already excited to
some extent. Later on, however, strong nonlinear effects develop, and
the final stationary pattern, Fig.~\ref{Fig2}(c), becomes very
different from the one determined by the critical mode. Observing the
nonlinear development (see Video~\cite{Video}), we notice that some
elements get progressively kicked off the main group near the
destabilized uniform solution in this process.  Eventually, in the
asymptotic stationary state, all elements become separated into two
groups. The correspondence between the indices and the degrees for the
considered scale-free network is displayed in Fig.~\ref{Fig2}(d). The
separation into two groups occurs only for the elements on the nodes
with relatively small degrees (roughly $i>200$, $k_{i}<24$), while the
elements on the nodes with high degrees ($i<200$, $k_{i}>24$) do not
separate.

\begin{figure}[ptbh]
\begin{center}
\includegraphics[width=1.0\hsize]{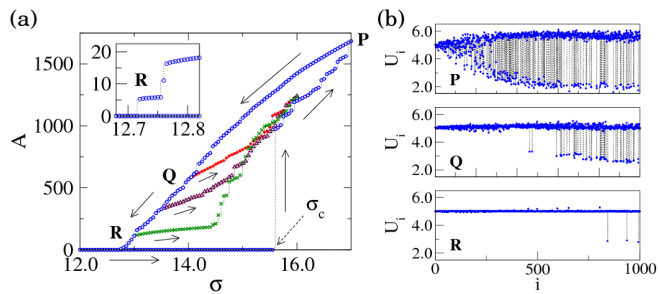}
\end{center}
\caption{(color online) (a) Amplitude $A$ of the Turing pattern
  vs. the diffusion ratio $\sigma$; variation direction of $\sigma$ is
  indicated by arrows. (b) Stationary Turing patterns at parameter
  points $P$ ($\sigma=17.0$), $Q$ ($\sigma=13.5$), and $R$
  ($\sigma=12.8$). The inset in (a) shows the blowup near $R$. }
\label{Fig3}
\end{figure}

The outcome of nonlinear evolution depends sensitively on initial
conditions. Different Turing patterns are possible at the same
parameter values and strong hysteresis effects have been observed. As
an example, Fig.~\ref{Fig3}(a) shows how the amplitude of the
stationary Turing pattern, defined as $A=\sum_{i=1}^{N}\left\{
  (U_{i}-U^{(0)})^{2}+(V_{i}-V^{(0)})^{2}\right\} $, varies under
gradual variation of the parameter $\sigma$ in the upward or downward
directions.  Stationary patterns observed at points $P$, $Q$, and $R$
in Fig.~\ref{Fig3}(a) are presented in Fig.~\ref{Fig3}(b). As $\sigma$
was increased starting from the uniform initial condition, the Turing
instability took place at $\sigma=\sigma_{c}$, with the amplitude $A$
suddenly jumping up to a high value that corresponds to the appearance
of a kicked-off group. If $\sigma$ was further increased, the
amplitude $A$ grew. Starting to decrease $\sigma$, we did not however
observe a drop down at $\sigma=\sigma_{c}$. Instead, a punctuated
decrease in the amplitude $A$, which is characterized by many
relatively small steps, was found. Reversing the direction of change
of the parameter $\sigma$ at different points, many coexisting
solution branches could be identified.  The characteristics of Turing
patterns vary with their amplitudes.  When $A$ is close to zero (point
$R$ in Fig.~\ref{Fig3}), only a few kicked-off elements remain in the
system.  Because of the hysteresis, solutions with only a small number
of destabilized elements can coexist with the linearly stable uniform
state in this region, so that the localized Turing patterns are found
for $\sigma < \sigma_{c}$.

The properties of the developed Turing patterns above the instability
boundary ($\sigma > \sigma_{c}$) can be relatively well understood
by using the mean-field approximation, previously used for the
analysis of epidemics spreading on networks~\cite{Pastor} and for
networks of phase oscillators~\cite{Ichinomiya}.  We start by writing
Eqs.~(\ref{Eq:RD}) in the form
\begin{align}
  \dot{U_{i}}(t) & =
  f(U_{i},V_{i})+\epsilon(h_{i}^{(U)}-k_{i}U_{i}),\cr
  \dot{V_{i}}(t) & = g(U_{i},V_{i})+\sigma\epsilon(h_{i}^{(V)}-k_{i}
  V_{i}),
\end{align}
where local fields felt by each element, $h_{i}^{(U)}=\sum_{j=1}^{N}
A_{ij}U_{j}$ and $h_{i}^{(V)}=\sum_{j=1}^{N}A_{ij}V_{j}$, are
introduced.  These local fields are further approximated as
$h_{i}^{(U)}\simeq k_{i} H^{(U)}$ and $h_{i}^{(V)}\simeq
k_{i}H^{(V)}$, where \emph{global mean fields} are defined by
$H^{(U)}=\sum_{j=1}^{N}w_{j}U_{j}$ and $H^{(V)}=\sum_{j=1}
^{N}w_{j}V_{j}$. The weights $w_{j}=k_{j}/\left(
  \sum_{j^{\prime}=1}^{N} k_{j}^{\prime}\right) =k_{j}/k_{total}$ take
into account the difference in the contributions of different nodes to
the global mean field, depending on their degrees
(cf.~\cite{Pastor,Ichinomiya}). Thus, the local fields are taken to be
proportional to the degree of a node, ignoring the details of its
actual connections.

With this approximation, each element interacts only with the global
mean fields, and its dynamics is described by
\begin{align}
  \label{Eq:Single}
  \dot{U}(t) &=f(U,V)+\beta(H^{(U)}-U),\cr
  \dot{V}(t)
  &=g(U,V)+\sigma\beta (H^{(V)}-V).
\end{align}
We have dropped here the index $i,$ since all elements obey the same
dynamics, and introduced the parameter $\beta=\epsilon k_{i}$. If
diffusion ratio $\sigma$ is fixed and the global mean fields $H^{(U)}$
and $H^{(V)}$ are given, this parameter $\beta$ plays the role of a
bifurcation parameter that controls the dynamics of each
element. Equations~(\ref{Eq:Single}) have a single stable fixed point
when $\beta=0$ (i.e. $\epsilon=0$), and, as $\beta$ is increased, this
system undergoes imperfect pitchfork bifurcations that give rise to
two new stable fixed points.

\begin{figure}[ptbh]
\begin{center}
\includegraphics[width=1.0\hsize]{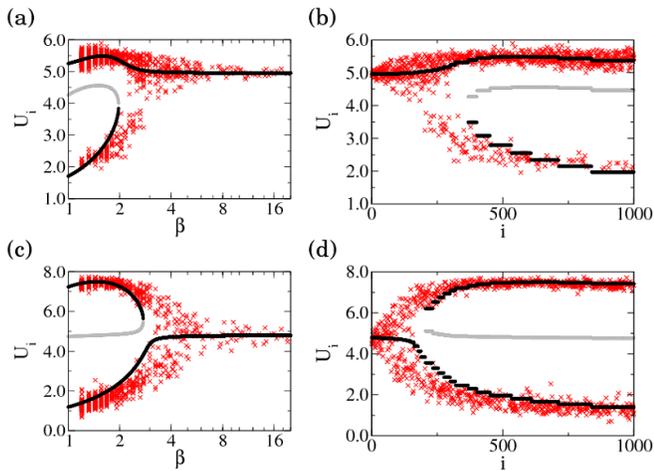}
\end{center}
\caption{(color online) Stationary Turing patterns compared with the
  bifurcation diagrams of a single element coupled to global mean
  fields. The parameters are $\epsilon=0.12$ and (a,b) $\sigma=15.6$,
  (c,d) $\sigma=30$. Black curves (dots) indicate stable branches,
  gray curves (dots) correspond to the unstable branch. Crosses show
  the computed Turing patterns.  The global mean fields are
  $(H^{(U)},H^{(V)})=(4.95,9.97)$ for $\sigma=15.6$, and
  $(H^{(U)},H^{(V)})=(4.8,9.9)$ for $\sigma=30$.}
\label{Fig4}
\end{figure}

We have computed stationary Turing patterns by numerical integration
of Eqs.~(\ref{Eq:RD}) and determined the respective global mean fields
$H^{(U)}$ and $H^{(V)}$ at $\sigma=15.6$ and $\sigma=30$. Substituting
these computed global mean fields into Eqs.~(\ref{Eq:Single}),
bifurcation diagrams for a single Mimura-Murray element have been
obtained (solid curves in Fig.~\ref{Fig4} (a,c)). These diagrams can
be compared with the actual stationary Turing patterns. Each node $i$
in the network is characterized by its degree $k_{i}$, so that it
possesses a certain value of the bifurcation parameter,
$\beta=\epsilon k_{i}$. Therefore, the Turing pattern can be projected
onto these bifurcation diagrams, as shown by crosses in
Fig.~\ref{Fig4}(a,c). We see a relatively good agreement between the
stable branches and the data from the actual Turing patterns.
Furthermore, we directly compare in Fig.~\ref{Fig4}(b,d) the computed
Turing patterns with the mean-field predictions, based on
Eqs.~(\ref{Eq:Single}). We see that the Turing patterns are nicely
fitted by the stable branches, though the scattering of numerical data
gets enhanced near the branching points. In this way, the fully
developed Turing patterns in our system are essentially explained by
the bifurcation diagrams of a single element coupled to constant
global mean fields, with the coupling strength determined by the
degree of the respective network node.

Thus, we have constructed a general linear theory of Turing patterns
in the activator-inhibitor systems on networks and have also performed
numerical investigations of nonlinear pattern formation.  The Turing
instability takes place on any networks as the ratio of diffusion
constants of the inhibitor and activator species is increased.  The
critical mode corresponds to a certain eigenvector of the Laplacian
matrix, but the final Turing patterns may be very different from the
critical mode; multistability of solutions and hysteresis are effects
are typically observed.  Localized Turing patterns below the linear
instability threshold have been found.

To understand this behavior, it should be taken into account that,
while being relatively large ($N=1000$), the considered random
networks had nonetheless only the small diameters of $4$ and,
moreover, each node in a network had $20$ neighbors on the average.
Therefore, diffusional mixing should be very strong in such
activator-inhibitor systems, and behavior characteristic of small
well-mixed spatial volumes can be expected; each element in a network
may be viewed as interacting with some global mean fields.  Indeed,
separation of elements into two clusters has previously been reported
for globally coupled activator-inhibitor systems under similar
conditions~\cite{Mizuguchi}.  The coupling to the global mean fields
is however essentially heterogeneous in our case, because it is
determined by the degrees of the respective network nodes which vary
greatly.  Although not included here, investigations have also been
undertaken by us for the classical Brusselator model and for the
Erd\"os-R\'enyi random networks, and similar results have been
obtained.  The constructed mean-field theory of nonlinear Turing
pattern formation is applicable for various large random networks of
small diameters and different activator-inhibitor systems.

Financial support of the Volkswagen Foundation (Germany) is gratefully
acknowledged.

\end{document}